\documentclass[twocolumn,aps,showpacs,showkeys, superscriptaddress]{revtex4}
\usepackage{graphicx}

\newcommand\beq{\begin{equation}}
\newcommand\eeq{\end{equation}}
\newcommand\bea{\begin{eqnarray}}
\newcommand\eea{\end{eqnarray}}

\newcommand{\bphi}{{\mbox{\boldmath $\varphi$}}}
\newcommand{\bone}{{\bf 1}}
\newcommand{\bC}{{\bf C}}
\newcommand{\cH}{{\cal H}}
\newcommand{\bL}{{\bf L}}
\newcommand{\bsin}{{\bf\sin}}
\newcommand{\bR}{{\bf R}}
\newcommand{\bM}{{\bf M}}
\newcommand{\bN}{{\bf N}}
\newcommand{\bPhi}{{\mbox{\boldmath $\Phi$}}}
\newcommand{\bS}{{\bf S}}
\newcommand{\bI}{{\bf I}}
\newcommand{\bF}{{\bf F}}
\newcommand{\bQ}{{\bf Q}}

\newcommand{\ba}{\begin{array}}
\newcommand{\ea}{\end{array}}

\begin{document}
\draft

\title{Efficient evaluation of decoherence rates in complex Josephson circuits}

\author{David~P.~DiVincenzo$ ^1$, Frederico Brito$ ^{2,1}$ and Roger H. Koch}

\affiliation{IBM T.J. Watson Research Center, P.O. Box 218,
Yorktown Heights, NY 10598 USA\\
$^2$Departamento de F\'isica da Mat\'eria Condensada,
Instituto de F\'isica Gleb Wataghin,
Universidade Estadual de Campinas, Campinas-SP 13083-970, Brazil}

\begin{abstract}
A complete analysis of the decoherence properties of a Josephson
junction qubit is presented.  The qubit is of the flux type and
consists of two large loops forming a gradiometer and one small
loop, and three Josephson junctions.  The contributions to
relaxation ($T_1$) and dephasing ($T_\phi$) arising from two
different control circuits, one coupled to the small loop and one
coupled to a large loop, is computed.  We use a complete,
quantitative description of the inductances and capacitances of
the circuit.  Including two stray capacitances makes the quantum
mechanical modeling of the system five dimensional.  We develop a
general Born-Oppenheimer approximation to reduce the effective
dimensionality in the calculation to one.  We explore $T_1$ and
$T_\phi$ along an optimal line in the space of applied fluxes;
along this ``S line" we see significant and rapidly varying
contributions to the decoherence parameters, primarily from the
circuit coupling to the large loop.
\end{abstract}


\maketitle

\section{Introduction}
\label{introduction}

Recent years have seen many successes in obtaining high-coherence
quantum behavior in a variety of flux-based Josephson-junction
qubits. The devices which show good behavior as qubits are fairly
complex electrical circuits, and a detailed theoretical analysis
of these circuits has proven useful in arriving at optimal designs
with the best decoherence behavior\cite{B1,B2}.  Since the first
reports of coherent oscillations in Josephson
qubits\cite{Nakamura}, the observed coherence times have increased
by a factor of about 5000; theory has had a substantial role in
this large increase (for a theoretical review of Josephson qubits,
see \cite{Makhlin}), by suggesting strategies for choosing optimal
settings of control parameters for the operation of the qubit.

In this paper, we report the results of a detailed theoretical
study of the flux qubit recently reported by our
group\cite{Kochetal}.  We will make extensive use of a method of
analysis introduced by Burkard, Koch, and DiVincenzo
(BKD)\cite{BKD}.  BKD introduced a universal method for analyzing
any electrical circuit that can be represented by lumped elements.
BKD proceeds in several steps: first, the Kirchhoff equations are
formulated in graph theoretic language so that they describe the
dynamics of a general circuit in terms of a set of independent,
canonical coordinates.  Then, one set of terms in these equations
of motion (the ``lossless" part) is seen to be generated by a
Hamiltonian describing a massive particle in a potential; the
number of space dimensions in which the particle moves is equal to
the number of canonical coordinates in the Kirchhof equations. The
``lossy" parts of the equations of motion are treated by
introducing a bath of harmonic oscillators, in the style of
Caldeira and Leggett\cite{CG}.

Finally the resulting total Hamiltonian, involving a system, a
bath, and a system-bath coupling, can be analyzed by standard
means to determine the decoherence parameters, $T_1$ and $T_\phi$,
of the first two eigenlevels of the system (the ``qubit").  $T_1$
is the energy loss rate of the qubit, while $T_\phi$, the ``pure
dephasing time", is related to the experimental parameter $T_2$,
the decay time of Ramsey fringes, by $T_2^{-1}={1\over
2}T_1^{-1}+T_\phi^{-1}$.  Long $T_1$ and $T_\phi$ times are both
necessary conditions for quantum computing.

\begin{figure}[htb]
\includegraphics[width=8cm,trim=0 0 0 0]{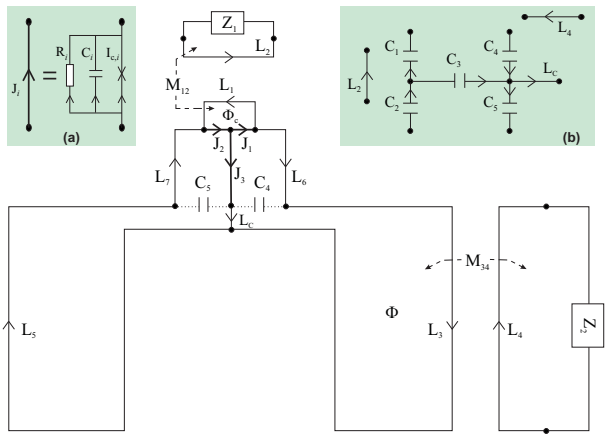}
\caption{The IBM qubit, drawn as a network graph.  This is an
oriented graph of the gradiometer structure coupled to two sources
$Z_1$ and $Z_2$ via mutual inductances $M_{12}$ and $M_{34}$.
Branches of the graph represent Josephson junctions $J_i$,
inductances $L_i$ and external impedances $Z_i$. Each Josephson
junction branch (thick line) is modelled by a resistively shunted
Josephson junction (RSJ) containing (inset (a)) an ideal junction
with critical current $I_{c,i}$, junction capacitance $C_i$ and
shunt resistance $R_i$.  $C_4$ and $C_5$ denote stray capacitances
present in the circuit. The qubit is operated changing the
external fluxes $\Phi_c$ and $\Phi$ applied through the small and
large loops, respectively. Inset (b): The tree chosen for the
graph. Values used for this qubit: $L\text{z}_{i}=Z_i(\omega)/i\omega$; $Z_1=Z_2=50\,\Omega,~I_c=\frac{\Phi_0}{2\pi L_J}=0.8\mu A,~C_i=10\text{f\,F},~ C_4=C_5=50\text{f\,F}$ and, using the modelling program FastHenry, $ L_c=106.27,~L_1=32.18,~L_3=L_5=605.03,~L_6=L_7=81.46,~L_2=32.18,~L_4=605.03,~M_{12}=0.8,~M_{13}=M_{15}=0.18,~M_{16}=M_{17}=-2.41,~M_{3c}=-M_{5c}=0.52,
~M_{34}=0.5,~M_{35}=3.4,~M_{36}=M_{57}=0.22,~M_{37}=M_{56}=-0.86,~M_{1c}=0 \text{ (exact),}~M_{6c}=-M_{7c}=27.63,~M_{67}=-13.93 \text{ (all in units of
pH).}$}
\label{qubit}
\end{figure}

The results of this paper have revealed significant facts about
the dependence of $T_1$ and $T_\phi$ on the control parameters of
our qubit.  The qubit has, to a good degree of approximation, a
bilateral symmetry across its midline (see Fig. \ref{qubit}). This
symmetry manifests itself in the quantum behavior: The quantum
structure is effectively that of a symmetric double well potential
whenever the difference of bias fluxes in the two large loops
$\Phi$ is the flux quantum $\Phi_0=h/2e$.  (The structure is a
``gradiometer", meaning that, to good approximation, its behavior
is only a function of the difference of the magnetic flux in the
two large loops.)
We will analyze the decoherence parameters arising from the two
impedances shown, $Z_1$ and $Z_2$. Because $Z_1$ is coupled to the
qubit via the ``small" loop, we refer to the decoherence
parameters associated with it as $T_{1s}$ and $T_{\phi s}$; the
corresponding parameters for $Z_2$, coupled via the ``large" loop,
are $T_{1l}$ and $T_{\phi l}$.   We find that the bilateral
symmetry completely controls the overall structure of the $T_1$s
and $T_\phi$s. All these parameters are symmetric in $\Phi$ around
$\Phi_0$ ($T_{\phi l}$ and $T_{1l}$ are approximately symmetric
for small values of control flux $\Phi<0.39\Phi_0$, the other two
are exactly symmetric). Furthermore, $T_{1s}$, $T_{\phi s}$, and
$T_{\phi l}$ all have divergent behavior at the symmetric point;
$T_{1s}$ is exactly divergent, $T_{\phi s}$ and $T_{\phi l}$ are
very nearly so for a large range of small-loop control flux
$\Phi_c$. These facts give a powerful motivation for operating the
qubit always very near $\Phi=\Phi_0$. As a function of $\Phi_c$,
$T_{\phi l}$ is strongly increasing and $T_{1l}$ is strongly
decreasing (in the symmetric situation). This makes it essential
to stay within a particular window of operating parameters.

As we will discuss in detail, the full dependence of the four
decoherence parameters on $\Phi$ and $\Phi_c$ is complex, but can
be grossly understood as being controlled by two distinct regimes,
the ``semiclassical" and the ``harmonic".  In the semiclassical
regime, the effective potential is a double well with a high
barrier between, so that quantum tunneling is very small.  As
$\Phi_c$ is increased, the barrier drops, then disappears
altogether; then the qubit potential enters the ``harmonic"
regime, where the potential is approximately just a single,
quadratic well.  These two extreme cases are relatively simple;
decoherence in the regime of crossover between these two is rather
complex.

We gain some further insight into these results via a technical
improvement that we have added to the analysis of BKD\cite{BKD}. A
description of this improvement, an application of the {\em
Born-Oppenheimer} approximation, is another important component of
this paper.  This improvement was motivated by the fact that we
wanted to study the effect of stray capacitances in the qubit
circuit of Fig. \ref{qubit}.  The quantum mechanics that this
model defines is that of a particle in a five-dimensional
potential (five because there are three junction capacitances and
two stray capacitances, each defining a degree of freedom).  A
direct evaluation of the Schroedinger equation in five dimensions
is numerically complex.  But we find that, in a controlled way, we
can organize these five dimensions into four coordinate directions
that are ``fast" (in which the potential rises very steeply) and
one that is ``slow" (and has the double-well structure at low
$\Phi_c$).  Then, just as in molecular physics \cite{Mertzbacher},
the fast coordinates can be treated adiabatically, having the
effect of modifying the effective slow potential energy in the one
remaining coordinate. The resulting one-dimensional quantum theory
is very easy to analyze numerically, and amenable to a qualitative
discussion.

This paper is organized as follows: Sec. II introduces the network
graph formalism that we use to analyze the quantum mechanics of
Josephson circuits.  We stress two innovations that considerably
streamline the analysis: a capacitance rescaling, and a
Born-Oppenheimer approximation. The Appendix gives more background
about the theory, with subsection A giving a review, with some
minor corrections, of the relevant parts of BKD\cite{BKD}, and
subsection B highlighting some new results in network graph
theory.  Sec. III discusses the details of the necessary
computation that are specific to the gradiometer qubit. Sec. IV
gives a qualitative discussion of the features of the four
decoherence parameters, $T_{1s}$, $T_{1l}$, $T_{\phi s}$, and
$T_{\phi l}$ that we compute.  Sec. V reviews a semiclassical
analysis from BKD\cite{BKD} that is helpful in understanding the
overall features of the decoherence parameters. The four Secs.
VI-IX give an extended discussion of each of the four decoherence
parameters. Sec. X gives some conclusions.

\section{Analysis: capacitance rescaling and Born Oppenheimer approximation}
\label{bo}

Our analysis follows closely that of BKD\cite{BKD}.  A summary of
the essentials of this theory is given in Appendix A.  The result
of this theory is, first, a system Hamiltonian, which we begin
with here (see Eq. (\ref{generalpotential})): \beq
\cH_S(t)={1\over
2}\bQ_C^T\bC^{-1}\bQ_C+\left({\Phi_0\over2\pi}\right)^2U(\bphi,t),
\eeq \bea
&&U(\bphi,t)=-\sum_iL_{J;i}^{-1}\cos\varphi_i\nonumber\\&&+{1\over
2}\bphi^T\bM_0\bphi+{2\pi\over\Phi_0}\bphi^T[(\bar{\bN}*\bPhi_x)(t)+
(\bar{\bS}*\bI_B)(t)]. \qquad\eea

To perform the Born Oppenheimer approximation, it is best to first
go to a rescaled coordinate system in which the mass (i.e., the
capacitance matrix $\bC$) is isotropic.  This is mentioned in
BKD\cite{BKD}, but we present this analysis more generally here to
set our notation. We make the following coordinate transformation
\begin{eqnarray}
&&{\bf q}=c^{1/2}\bC^{-1/2}\bQ_C,\label{rescaledmomentum}\\
&&{\bf f}=c^{-1/2}\bC^{1/2}\bphi,\label{rescaledphase}
\end{eqnarray}
$c$ is some standard capacitance; it is convenient to insert this
arbitrary number so that $q$ and $\text{f}$ have the same units as $Q_C$
and $\varphi$, respectively.  Note that the commutation relations
are left unchanged by this coordinate change:
\begin{equation}
{\Phi_0\over
2\pi}(\varphi_iQ_{C,j}-Q_{C,j}\varphi_i)=i\hbar\delta_{ij}\,\rightarrow\,{\Phi_0\over
2\pi}(\text{f}_iq_j-q_j\text{f}_i)=i\hbar\delta_{ij}.
\end{equation}
The Hamiltonian for the rescaled Schroedinger equation is \beq
\cH_S(t)={1\over 2c}{\bf q}^T{\bf q}+\left({\Phi_0\over
2\pi}\right)U'({\bf f},t), \eeq
\begin{widetext}
\bea \lefteqn{U'({\bf
f},t)=-\sum_iL_{J;i}^{-1}\cos(c^{1/2}(\bC^{-1/2}{\bf
f})_i)}\nonumber\\&&\qquad\quad\quad\,\:+{1\over 2}{\bf
f}^T(c\bC^{-1/2}\bM_0\bC^{-1/2}){\bf f}+{2\pi\over\Phi_0}{\bf
f}^T[c^{1/2}\bC^{-1/2}(\bar{\bN}*\bPhi_x)(t)+
c^{1/2}\bC^{-1/2}(\bar{\bS}*\bI_B)(t)]. \label{rescaledpotential}\eea
\end{widetext}
For computing decoherence parameters, we take over unchanged the
golden-rule formulas discussed in BKD\cite{BKD} (see Appendix):
\begin{eqnarray}
  \frac{1}{T_1} &=& 4|\langle 0|{\bf m}\cdot\bphi|1\rangle|^2 J(\omega_{01}) \coth\frac{\omega_{01}}{2k_B T}, \label{T1}\\
  \frac{1}{T_\phi} &=&  |\langle 0|{\bf m}\cdot\bphi|0\rangle-\langle 1|{\bf m}\cdot\bphi|1\rangle|^2 \left.\frac{J(\omega)}{\omega}\right|_{\omega\rightarrow 0} \!\!\!\!\!\!\!\!\! 2k_B T. \quad\quad\label{Tphi}
\end{eqnarray}
For the rescaled coordinates, these are
\begin{widetext}
\begin{eqnarray}
 \frac{1}{T_1} &=& 4|\langle 0|c^{1/2}{\bf m}^T\bC^{-1/2}{\bf f}|1\rangle|^2 J(\omega_{01}) \coth\frac{\omega_{01}}{2k_B T}, \label{T1}\\
  \frac{1}{T_\phi} &=&  |\langle 0|c^{1/2}{\bf m}^T\bC^{-1/2}{\bf f}|0\rangle-\langle 1|c^{1/2}{\bf m}^T\bC^{-1/2}{\bf f}|1\rangle|^2 \left.\frac{J(\omega)}{\omega}\right|_{\omega\rightarrow 0} \!\!\!\!\!\!\!\!\! 2k_B T. \quad\quad\label{Tphi}
\end{eqnarray}
\end{widetext}
We will discuss the use of the Born-Oppenheimer approximation to
evaluate these formulas.  What must be computed are matrix
elements of the form
\begin{equation}
\int d{\bf f}({\bf v}\cdot{\bf f})\langle
\alpha|{\bf f}\rangle\langle{\bf f}|\beta\rangle,
\label{matrixelement}
\end{equation}
where $\alpha,\beta=0,1$, and $\bf v$ is the constant vector
$c^{1/2}{\bf m}^T\bC^{-1/2}$.

As discussed in the introduction, we single out one (more than one
is also possible) ``slow" degree of freedom $\rm{f}_\|$, and take
all coordinate directions orthogonal to this one, $\bf{f}_\bot$,
to be ``fast".  So
\begin{equation}
{\bf f}= \{ {\rm f}_\|,{\bf f}_\bot \}.
\end{equation}
The fast coordinates are characterized by the fact that the
potential $U'({\bf f})$ increases very rapidly in the ${\bf
f}_\bot$-direction; we assume that it is a good approximation to
expand in these directions to second order:
\begin{equation} U'({\bf f})\approx
V({\rm f}_\|)+\sum a_i({\rm f}_\|){\bf f}_{\bot,\,i}+\sum
b_{ij}({\rm f}_\|){\bf f}_{\bot,\,i}{\bf f}_{\bot,\,j},
\label{quadraticpotential}
\end{equation}
where {\bf b} can be taken to be a real symmetric matrix.

In this case, the Born-Oppenheimer approximation is made as
follows\cite{Mertzbacher}: fix the slow coordinate ${\rm f}_\|$,
solve the remaining (harmonic) Schroedinger equation in fast
coordinates ${\bf f}_\bot$.  The ground state eigenvalue of this
Schroedinger equation is
\begin{equation}
u({\rm f}_\|)=\left({\Phi_0\over
2\pi}\right)^2V({\rm f}_\|)-\left({\Phi_0\over
2\pi}\right)^2{1\over 4}\,{\bf a}^T{\bf b}^{-1}{\bf
a}+{\hbar\over\sqrt{2c}}\,{\rm Tr}\,\sqrt{{\bf b}}.
\end{equation}
Note that this effective potential has nontrivial ${\rm f}_\|$
dependence from its last two terms. The first and second terms
represent the value of the potential (in the ${\bf f}_\bot$
coordinates), and the final term is the sum of the zero point
energies ${1\over 2}\hbar\omega$ in this multidimensional harmonic
well.

The minimum of the potential in the ${\bf f}_\bot$ coordinates, as
a function of ${\rm f}_\|$, is
\begin{equation}
{\bf f}_\bot^{min}({\rm f}_\|)=-{1\over 2}{\bf
b}^{-1}({\rm f}_\|){\bf a}({\rm f}_\|).
\end{equation}
The ground state wavefunction in the ${\bf f}_\bot$ coordinates is a
gaussian centered at this point, which we will indicate as
\begin{equation}
\langle{\bf f}_\bot|\,0,{\rm f}_\|\rangle=g({\bf f}_\bot-{\bf f}_\bot^{min}({\rm f}_\|)).
\end{equation}
In the Born-Oppenheimer approximation, the full wavefunction is
taken to be
\begin{equation}
\langle{\bf f}|\alpha\rangle=
\langle{\rm f}_\||\alpha\rangle\langle{\bf f}_\bot|\,0,{\rm f}_\|\rangle=
\langle{\rm f}_\||\alpha\rangle
g({\bf f}_\bot-{\bf f}_\bot^{min}({\rm f}_\|)).
\end{equation}
Where $\langle{\rm f}_\||\alpha\rangle$ is the $\alpha^{th}$
eigenstate of the one-dimensional, slow-coordinate Schroedinger
equation
\begin{equation}
\left[-\left({2\pi\over\Phi_0}\right)^2{\hbar^2\over 2c}{d^2\over
d{\rm f}_\|^2}+u({\rm f}_\|)\right]\langle{\rm f}_\||\alpha\rangle
=\lambda_\alpha\langle{\rm f}_\||\alpha\rangle.\label{oneD}
\end{equation}
We return to the matrix elements that are to be computed, Eq. (\ref{matrixelement}). We separate the integrand into a fast and a slow part:
\begin{eqnarray}
\label{fastslow} \int&&\!\!\!\!d{\bf f}({\bf v}\cdot{\bf
f})\langle \alpha|{\bf f}\rangle\langle{\bf f}|\beta\rangle=\int
d{\bf f}({\bf v_\bot}\cdot{\bf f}_\bot+{\rm v}_\|{\rm
f}_\|)\langle
\alpha|{\bf f}\rangle\langle{\bf f}|\beta\rangle\nonumber\\
=&&\!\!\!\!\!\!\int d{\rm f}_\|d{\bf f}_\bot {\bf
v_\bot}\cdot{\bf f}_\bot\langle\alpha|{\rm f}_\|\rangle\langle{\rm f}_\||\beta\rangle
g^2({\bf f}_\bot-{\bf f}_\bot^{min}({\rm f}_\|))+\nonumber\\&&\ \int
d{\rm f}_\|d{\bf f}_\bot
{\rm v}_\|{\rm f}_\|\langle\alpha|{\rm f}_\|\rangle\langle{\rm f}_\||\beta\rangle
g^2({\bf f}_\bot-{\bf f}_\bot^{min}({\rm f}_\|))\nonumber\\=&&\!\!\!\!\!\!\int
d{\rm f}_\|\langle\alpha|{\rm f}_\|\rangle\langle{\rm f}_\||\beta\rangle\int
d{\bf f}_\bot{\bf v_\bot}\cdot{\bf f}_\bot
g^2({\bf f}_\bot-{\bf f}_\bot^{min}({\rm f}_\|))+\nonumber\\&&\ \int
d{\rm f}_\|{\rm v}_\|{\rm f}_\|\langle\alpha|{\rm f}_\|\rangle\langle{\rm f}_\||\beta\rangle\int
d{\bf f}_\bot
g^2({\bf f}_\bot-{\bf f}_\bot^{min}({\rm f}_\|))\nonumber\\
=&&\!\!\!\!\!\!\int d{\rm f}_\|({\bf
v_\bot}\cdot{\bf f}_\bot^{min}({\rm f}_\|))\langle\alpha|{\rm f}_\|\rangle\langle{\rm f}_\||\beta\rangle+\nonumber\\&&\ \int
d{\rm f}_\|{\rm v}_\|{\rm f}_\|\langle\alpha|{\rm f}_\|\rangle\langle{\rm f}_\||\beta\rangle.
\label{big}
\end{eqnarray}
In the last line we use the fact that the gaussian is a normalized
transverse wavefunction.  The final two-term expression of Eq.
(\ref{big}) will be used below in the evaluation of the $T_1$ and
$T_\phi$ expressions.  The Schroedinger equation solutions (Eq.
(\ref{oneD})) and all the necessary integrations are performed
numerically in Mathematica.

\section{results for the gradiometer qubit}
\label{results}

We have calculated the coherence properties of the gradiometer
qubit of Koch {\em et al.}\cite{Kochetal}, assuming coupling to
two different lossy circuits, one inductively coupled to the small
loop, and the other inductively coupled to one of the large loops
(see Fig. \ref{qubit}). Here we do not include the additional structure
considered in \cite{Kochetal}, a low-loss terminated transmission
line inductively coupled to the other large loop (not shown). This
structure strongly modifies the quantum behavior of the qubit when
the energy splitting of the ground and first excited state of the
qubit is large (comparable to 1.5GHz, a typical resonant frequency
for the terminated transmission line); however, for smaller energy
gaps this structure is expected to be unimportant.  The two lossy
structures included are expected to account for most of the
dissipative and decohering processes seen by the qubit.

It is known that the decohering effect of two such structures is
non-additive, see Brito and Burkard (Ref. \cite{BB}); but they show that
this nonadditive effect is typically small, and we will consider
the irreversible effects of each structure separately.

We have extended the analysis of \cite{Kochetal} to include the
effect of stray capacitances on the qubit quantum behavior. We
approximate the distributed stray capacitances as two new lumped
circuit elements, shown with dotted lines in Fig. \ref{qubit}.  Including
these capacitors, the circuit theory leads to a quantum
description of the qubit that is equivalent to that of a particle
in a five-dimensional potential.  Using the Born-Oppenheimer
analysis developed in this paper, the complexity of the
calculation is not too greatly increased by these additional
capacitances.  As we will see, these extra capacitances, even
though their capacitances are larger than the junction
capacitances, cause only quantitative differences in the behavior
of the decoherence parameters.
\begin{figure}[!t]
\includegraphics[width=8cm,trim=0 0 0 0]{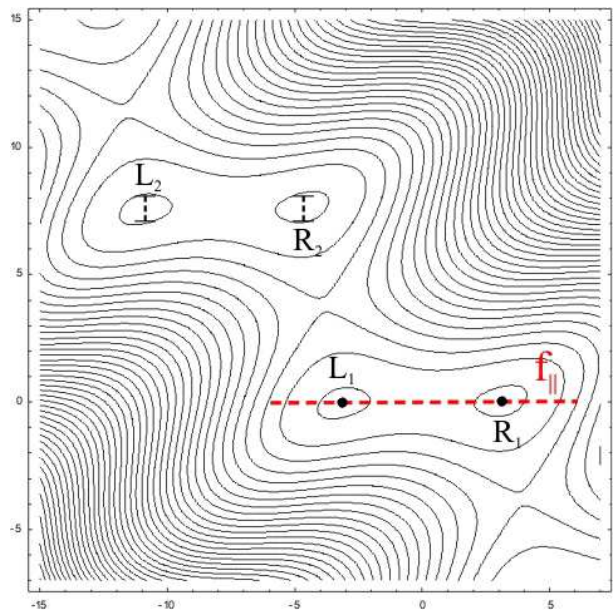}
\caption{Contour plot of the potential $U'({\bf f})$ on the S line
for the external fluxes $\Phi_c=0.36\Phi_0$ and $\Phi=\Phi_0$. The
red dashed line indicates the ``slow" direction ${\rm f}_\|$.
Along this direction the potential is a symmetric double well,
with the two relevant minima of the potential indicated by dots.
The bars show the spatial extension of the wave function, in the
vicinity of the minima, in the ``fast" direction ${\rm f}_\bot$
with the smallest curvature of the potential.}
\label{contourplot}
\end{figure}

Fig. \ref{contourplot} shows a two-dimensional slice of the
potential $U'$, after rescaling the capacitance matrix as
indicated in Eqs. (\ref{rescaledmomentum}, \ref{rescaledphase}).
The slice is chosen to include the two eigendirections of the
rescaled curvature matrix of the quadratic part of the $U'$
potential ($\bC^{-1/2}\bM_0\bC^{-1/2}$ of Eq.
\ref{rescaledpotential}).  In one of these directions the
curvature is zero; in this direction only the Josephson energy is
nonzero, and the potential is periodic (about two periods are
shown in the figure).  This periodicity reflects the $2\pi$
periodicity of the superconducting phase of the central island of
the circuit (the place where $J_1$, $J_2$, and $J_3$ meet in Fig.
\ref{qubit}). The displacement of the two-dimensional plane shown
in Fig. \ref{contourplot} is chosen so that the inductive energy
is minimized --- recall that the inductive energy consists of a
quadratic and a linear part.
\begin{figure}[!t]
\includegraphics[width=8cm,trim=0 0 0 0]{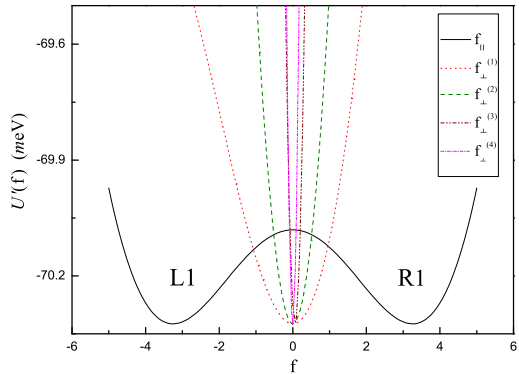}
\caption{Plots of the potential $U'$, in the vicinity of the
relevant minima, for each one of the orthogonal components of {\bf
f}. These plots were done by keeping fixed four coordinates at
their minimum points and varying the other one. For the ``slow"
direction ${\rm f}_\|$ (solid line) we see a double well structure
(symmetric on S line).  Along the ``fast" coordinates ${\rm
f}_{\bot}^{(i)}$, here calculated at the L1 point, a an almost
harmonic well is present. The external fluxes used for these plots
were $\Phi_c=0.36\Phi_0$ and $\Phi=\Phi_0$.}
\label{potential}
\end{figure}
The two dots in Fig. \ref{contourplot} indicate the minimum energy points of the
potential in this plane, which is almost (but not precisely) the
position of the absolute minima (these have also a small component
in the other three coordinate directions).  We choose the ``slow"
coordinate ${\rm f}_\|$ of the Born-Oppenheimer approximation to be
along the line connecting the two minima in the plane shown; the
other four directions are treated as the ``fast coordinates".

Fig. \ref{potential} gives more detail about the potential in
these ``fast" directions.  As expected, the potential rises more
steeply in all these directions than in the ``slow" direction. The
potentials are all basically harmonic, with some noticeable
anharmonicity, particularly in the softest ``fast" direction ${\rm
f}_\bot^{(1)}$.   But a calculation of the extent of the ground
wavefunction in this direction (error bars near L2 and R2 in Fig.
\ref{contourplot}) shows that it remains well confined within the
harmonic region.

\begin{figure}[!t]
\includegraphics[width=8cm,trim=0 0 0 0]{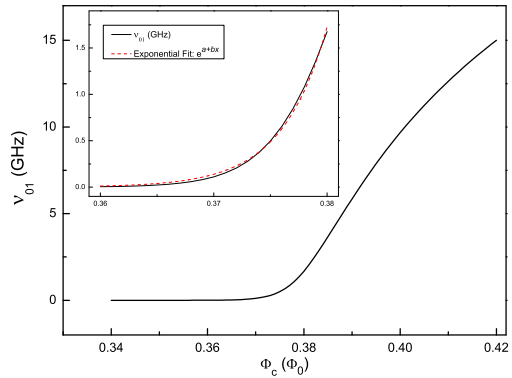}
\caption{The energy difference between the ground and first
eigenstate, $h \nu_{01}=\langle 1|{\cal H}_S|1\rangle-\langle
0|{\cal H}_S|0\rangle$, as a function of the ``control" flux
$\Phi_c$ on S line. Inset: Detailed view.  The red dashed line
represents a exponential fit of the data, giving a reasonable
representation of the data in this vicinity.}
\label{f01inset}
\end{figure}

\begin{figure}[!t]
\vspace*{-0.3cm}\hspace*{-0.7cm}\includegraphics[width=10cm,trim=0 0 0 0]{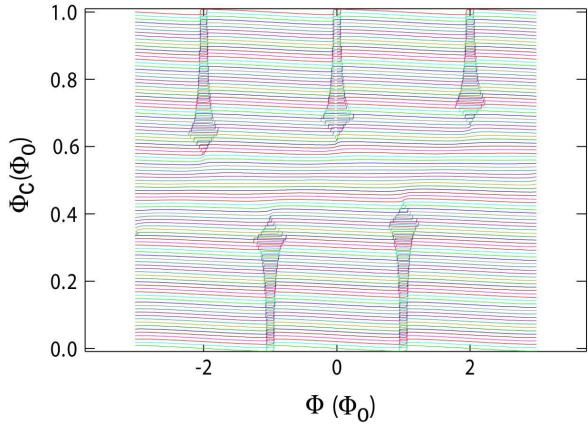}
\vspace*{-8ex}\caption{Simulated scans of the critical current over a wide
region of the $\Phi$-$\Phi_c$ plane.  These scans periodically
show hysteresis in vertically oriented regions in this plane,
indicating the presence of the double minimum of the potential in
these regions.}
\label{davefig}
\end{figure}

\begin{figure}[!t]
\includegraphics[width=8cm,trim=0 0 0 0]{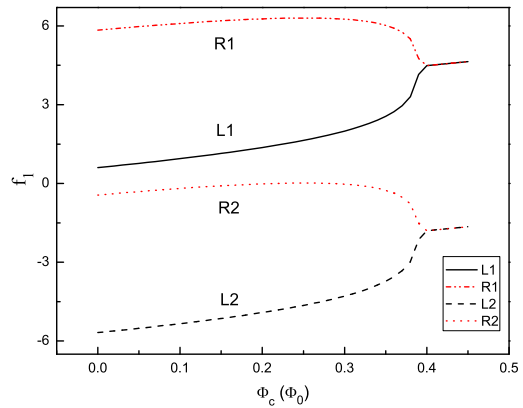}
\caption{The values of the phase associated with the Josephson
junction $J_1$ at the minima of the potential, as function of the
``control" flux $\Phi_c$, along the S line. Two consecutive pairs
of minima (L$i$-R$i$) along the periodic direction are shown. Near
$\Phi_c\gtrsim 0.39$, the double-minimum structure collapses
rapidly to a single minimum per period; in this regime the
distance between the single minima ($i$)-($i+1$) is $2\pi$. }
\label{phase1}
\end{figure}

\begin{figure}[!t]
\includegraphics[width=8cm,trim=0 0 0 0]{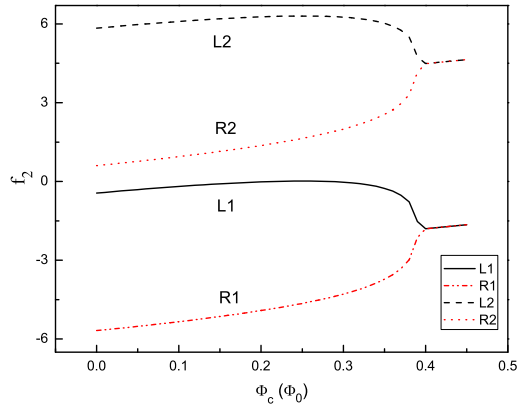}
\caption{The values of the phase associated with the
Josephson junction $J_2$ at the minima of the potential, as
function of the ``control" flux $\Phi_c$, along the S line.  Two
consecutive pairs of minima (L$i$-R$i$) along the periodic
direction are shown. Near $\Phi_c\gtrsim 0.39$, the double-minimum
structure collapses rapidly to a single minimum per period; in
this regime the distance between the single minima ($i$)-($i+1$)
is $2\pi$.}
\label{phase2}
\end{figure}

\begin{figure}[!t]
\includegraphics[width=8cm,trim=0 0 0 0]{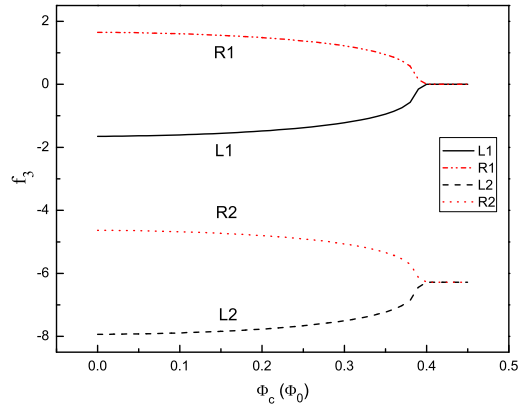}
\caption{The values of the phase associated with the Josephson
junction $J_3$ at the minima of the potential, as function of the
``control" flux $\Phi_c$, along the S line. Two consecutive pairs
of minima (L$i$-R$i$) along the periodic direction are shown. Near
$\Phi_c\gtrsim 0.39$, the double-minimum structure collapses
rapidly to a single minimum per period; in this regime the
distance between the single minima ($i$)-($i+1$) is $2\pi$.}
\label{phase3}
\end{figure}

\begin{figure}[!t]
\includegraphics[width=8cm,trim=0 0 0 0]{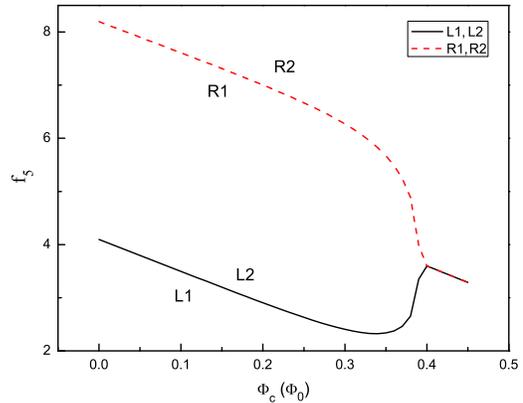}
\caption{The values of the phase associated with the stray
capacitance $C_5$ at the minima of the potential, as function of
the ``control" flux $\Phi_c$, along the S line.  (The phase of a
capacitance is proportional to the time integral of the voltage
across the capacitor.)  For this phase all minima pairs
(L$i$-R$i$) with the same values. This occurs because of the {\it
absence} of a Josephson energy term dependent on this phase.}
\label{phase5}
\end{figure}

\begin{figure}[!t]
\includegraphics[width=8cm,trim=0 0 0 0]{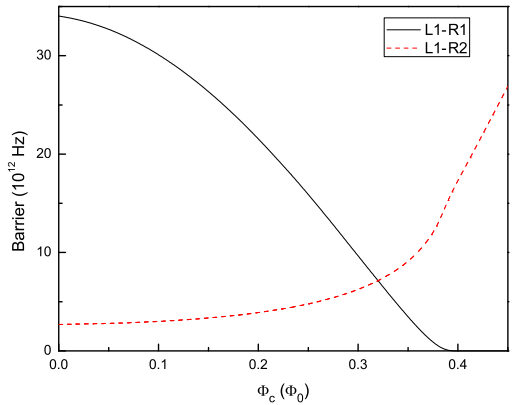}
\caption{The value of the potential barrier on the S line as a
function of $\Phi_c$.  Both the L1-R1 barrier height (solid black)
and the L1-R2 barrier height (dashed red) are shown.  The height
of the barrier separating the L1-R1 decreases rapidly as these
minima approach one another. For $\Phi_c\gtrsim 0.39$, the barrier
vanishes entirely between L1 and R1.}
\label{barrier}
\end{figure}

We have chosen a ``symmetric" setting for the parameters,
$\Phi=\Phi_0$, such that the potential is a symmetric double well
--- the depth of the pair of potential minima in Fig. \ref{contourplot} is equal.
This defines a line in the $\Phi$-$\Phi_c$ plane that we refer to
as the ``S line" (S for symmetric).  As the external control
parameters $\Phi$ and $\Phi_c$ are varied, this potential
landscape is changed in two different ways:
\begin{enumerate}
\item As $\Phi_c$ is varied, the distance between the two minima,
and thus the height of the barrier separating them, varies.
Increasing $\Phi_c$ from the value shown, $\Phi_c=0.36\Phi_0$, the
distance between L1 and R1 (L and R for ``left" and ''right")
drops rapidly, as shown by Figs. \ref{phase1}-\ref{phase5}, which
show how these minimum points evolve as a function of $\Phi_c$
along the S line.  As the minima approach one another, the height
of the barrier separating the L1 and R1 minima decreases rapidly,
as shown in Fig. \ref{barrier}. In this regime the
quantum-mechanical tunnel splitting between the lowest-lying
energy levels increases dramatically, see Fig. \ref{f01inset}.
Around $\Phi_c=0.39\Phi_0$ the barrier vanishes entirely. There
follows a long interval of $\Phi_c$ in which there is only a
single minimum per period of the potential; when $\Phi_c$
increases a little beyond $\Phi_c=0.39\Phi_0$, the potential
becones quite harmonic around its minimum. \item As $\Phi$ is
varied around $\Phi_0$, the energies of the two minima are shifted
with respect to one another.  For larger excursions of $\Phi$ away
from $\Phi_0$, one minimum becomes unstable, and only one minimum
per period remains stable.
\end{enumerate}
Fig. \ref{davefig} shows a large region of the $\Phi$-$\Phi_c$
plane, simulating a sequence of measurements very much as they are
done in the experiment: For a sequence of values of $\Phi_c$,
$\Phi$ is scanned from left to right and back again.  Each scan
(nearly horizontal line) plots the value of $\Phi_c$, plus a signal
proportional to the classical circulating current in one of the
large loops of the qubit.  The most prominent feature of this
sequence of curves is the thin vertical regions in which the scans
are hysteretic.  This essentially plots the region in which there
is a double minimum in the potential.  The shape of this region
reflects the behavior of the barrier height with control flux,
Fig. \ref{barrier}.  Looking at flux $\Phi=\Phi_0$, one sees, as
one decreases the control flux $\Phi_c$ from about 0.4$\Phi_0$, a
rapid widening of the hysteresis feature, reflecting a rapid
increase of the barrier height.  The abrupt switch to shrinkage of
the hysteresis loop reflects a switching of the lowest barrier
from the L1-R1 line to the L1-R2 line.  This cuspy feature is
readily seen in the experiment \cite{Kochnew}, and is an excellent
landmark for calibrating the actual applied fluxes.
Fig. \ref{davefig} is clearly periodic with changes in applied
flux.  Since $\Phi$ and $\Phi_c$ are Aharonov-Bohm fluxes (i.e.,
involving no magnetic field penetrating the interior of the
conductors), changing either by an integer multiple of $\Phi_0$
should leave the quantum behavior of the system invariant.  This
is actually {\em not} the periodicity that is seen in Fig.
\ref{davefig}. This absence of Aharonov-Bohm periodicity, an
apparent violation of gauge invariance, is a result of the fact
that the outer perimeter of the qubit is not interrupted by a
Josephson junction; because the temperature is very low compared
with the superconducting energy gap, there is a very high barrier
to the motion of a flux quantum into or out of the device. If this
barrier is assumed to be infinite (as it effectively is in our
model), the states of the device fall into noncommunicating
sectors.

Within these sectors, there remains the periodicity with respect
to varying the external fluxes seen in the figure: we can show
that if $\Phi$ is changed by an integer multiple of $\Phi_0$,
$(k_1-k_2-2k_3)\Phi_0$ (Each $k_i$ is any integer), the qubit
Hamiltonian is invariant if $\Phi_c$ is simultaneously changed by
$-(k_1+k_2)-\frac{L_1}{L_3+L_6}(k_2+k_3)$.  This shift of $\Phi$
and $\Phi_c$ are associated with the phase changes $\Delta {\rm
f}_1=2\pi k_1,\Delta{\rm f}_2=2\pi k_2,\Delta {\rm f}_3=2\pi k_3$,
$\Delta {\rm f}_{4}=
\frac{2\pi}{\sqrt{c}}\big(-k_1+k_3+\frac{L_6}{L_3+L_6}(k_2+k_3)\big)$
and $\Delta {\rm f}_{5}=
-\frac{2\pi}{\sqrt{c}}\frac{L_3}{L_3+L_6}(k_2+k_3)$. The
inductance factors in these expressions are approximate: they are
only true in the limit that all mutual inductances are zero. The
pattern of invariance as described by these equations is closely
matched in experimental data.

The construction of the quadratic and linear parts of the
potential in Eq. (\ref{rescaledpotential}) require a
graph-theoretic analysis of the gradiometer circuit, Fig.
\ref{qubit}.  An appropriate tree for the circuit graph is shown
in the inset (b) of Fig. \ref{qubit}.  Using this, the loop
matrices defined in the Appendix, Eq. (\ref{loopmatrix}), can be
read off by inspection:
\begin{equation}
F_{CL}=\left(\begin{array}{rrrrr}-1&0&0&-1&0\\-1&0&0&0&-1\\
0&0&0&1&-1\\0&1&0&-1&0\\0&0&1&0&-1\end{array}\right),\,\,
F_{CZ}=\left(\begin{array}{rr}0&0\\0&0\\
0&0\\0&0\\0&0\end{array}\right),
\end{equation}
\begin{equation}
F_{KL}=\left(\begin{array}{rrrrr}0&0&0&0&0\\0&0&0&0&0\\
0&1&-1&0&0\end{array}\right),\,\,
F_{KZ}=\left(\begin{array}{rr}-1&0\\0&-1\\
0&0\end{array}\right).
\end{equation}

For the numerical analysis of decoherence parameters, we need
values for the physical parameters of the circuit.  For the {\bf
C} matrix, circuit modeling indicates that we can take it to be a
diagonal matrix with diagonal elements $\{10,10,10,50,50\}$ (in
units of fF).  The 10fF capacitances are for the Josephson
junctions, the 50fF capacitances are the ``strays".  Although the
strays are numerically the largest capacitances, they do not
affect the results qualitatively, because of their positions in
the circuit.

The {\bf L} matrices are denoted:
\begin{equation}
L=\left(
\begin{array}{lllll}
L_1 & M_{13} & M_{15} & M_{16} & M_{17} \\
 M_{13} & L_{3} & M_{35} & M_{36} & M_{37} \\
 M_{15} & M_{35} & L_{3} & M_{56} & M_{57} \\
 M_{16} & M_{36} & M_{56} & L_6 & M_{67} \\
 M_{17} & M_{37} & M_{57} & M_{67} & L_6
\end{array}
\right),
\end{equation}
\begin{equation}
L_{LK}=\left(
\begin{array}{lll}
 M_{12} & 0 & M_{1c} \\
 0 & M_{34} & M_{3c} \\
 0 & 0 & M_{5c} \\
 0 & 0 & M_{6c} \\
 0 & 0 & M_{7c}
\end{array}
\right),
\end{equation}

\begin{equation}
L_{K}=\left(
\begin{array}{lll}
 L_2 & 0 & 0 \\
 0 & L_4 & 0 \\
 0 & 0 & L_{c}
\end{array}
\right),\,\,
L_{Z}=\left(
\begin{array}{ll}
 L\text{z}_{1} & 0 \\
 0 & L\text{z}_{2}
\end{array}
\right).
\end{equation}
The numerical values of these parameters are given in the caption
of Fig. \ref{qubit}.

The decoherence parameters involve the temperature, which we take
as $T=5K$.  This rather high temperature, much larger than the
bath temperature of a dilution refrigerator, is an accurate
reflection of the effective noise temperature of the circuits
coupled to the qubit. Future experiments are planned which will
make this effective temperature much lower.

The formal applied flux vector is, ${\bf
\Phi_x}=\{\Phi_c,\Phi,\Phi_{\rm p},0,0\}$. $\Phi_c$ and $\Phi$
have been introduced previously, and $\Phi_{\rm p}$ is the flux in
the third loop, the {\em pick up loop}.  $\Phi_{\rm p}$ will
always to be taken to be zero in the analyses here.

With these matrices we compute the coefficients $\bM_0$ and
$\bar\bN(\omega)$ using the formulas in the Appendix (Eqs.
\ref{M0}, \ref{Nbar}) ($\bar\bS(\omega)$ does not occur, as no
current sources are present in the circuit).  The applied fluxes
are time-dependent in the experiments that we are modelling, so in
principle we need to retain the full frequency dependence of
$\bar\bN(\omega)$.  The presence of a frequency dependence in this
operator is indicative of a retardation phenomenon: the
Hamiltonian at time $t$ is {\em not} a function only of the
applied fluxes at time $t$; rather, because of the lossy elements
in the circuit, $H(t)$ depends on a convolution of $\bPhi_x$ over
times preceding $t$.  We find, however, that the range in time of
the kernel $\bar\bN(t)$ in this convolution is very short: this
time range is set by $L_2/{\rm Re}(Z_1(\omega=0))$ and $L_4/{\rm
Re}(Z_2(\omega=0))$.  For our parameters, this time is no more
that 10 psec. In experiments\cite{Kochetal}, the applied fluxes
are varied on a time scale greater than 100psec. For this reason,
we ignore this retardation effect in all our calculations here,
and set $\bar\bN(\omega)=\bar\bN(\omega=0)$.

\section{discussion of $T_1$ and $T_\phi$}
\label{discussion}

Figures \ref{t1s}, \ref{t1l}, \ref{tphs} and \ref{tphl} show the obtained dissipation and
decoherence rates obtained for the gradiometer qubit in the
vicinity of the symmetric line, shown as a function of changes in
the small- and large-loop bias fluxes ($\Phi_{\rm p}$ is taken to be zero
throughout).  The dependences of these quantities is complex, with
variations over a large range of values (note that all the plots
are logarithmic).  We can explain all the trends seen in these
curves.  Several key facts determine the overall structure of
these curves:\begin{figure}[!t]
\includegraphics[width=8cm,trim=0 0 0 0]{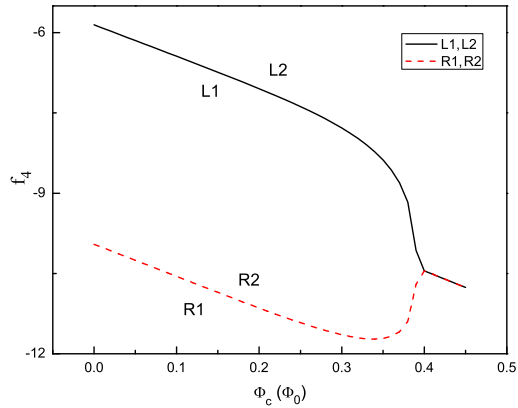}
\caption{The values of the phase associated with the stray
capacitance $C_4$ at the minima of the potential, as function of
the ``control" flux $\Phi_c$, along the S line.  (The phase of a
capacitance is proportional to the time integral of the voltage
across the capacitor.)  For this phase all minima pairs
(L$i$-R$i$) with the same values. This occurs because of the {\it
absence} of a Josephson energy term dependent on this phase.}
\label{phase4}
\end{figure}\begin{itemize}
\item Many of the curves have a break around $\Phi=(1\pm
\delta)\Phi_0$, $\delta\approx 0.01$. This is a consequence of a
level crossing that occurs near this value of $\delta$: for larger
$|\delta|$ the lowest two energy eigenvalues of the qubit are both
in one energy well. Thus, for $|\delta|>0.01$ the system is too
unsymmetrical for the two qubit states to correspond to the left
and right wells, and consequently the results in this regime are
not of great interest to us. \item For small values of the control
flux $\Phi_c\lesssim0.39\Phi_0$ the barrier is high, and the wave
function weight is concentrated near the minima of the two wells.
In this regime, which was referred to as the ``semiclassical"
regime in BKD, the various curves vary in predictable ways as the
barrier height and well asymmetry are changed, as we will detail
shortly. \item For large values of the control flux
$\Phi_c\gtrsim0.39\Phi_0$ the barrier vanishes, and the single
remaining well rapidly becomes almost exactly harmonic.  It is
straightforward to calculate what happens to $T_{1}$ and
$T_{\phi}$ in this harmonic limit, and we will see that the
data in this regime can be understood with reference to this
limit. \item The lossy circuit coupled to the small loop respects
the bilateral symmetry of the gradiometer qubit.  An exact
consequence of this is that $T_{1s}$ and $T_{\phi s}$ are
mathematically symmetric around $\Phi=\Phi_0$.  \item The lossy
circuit coupled to the large loop does {\em not} respect the
bilateral symmetry of the qubit.  Consequently, $T_{1l}$ and
$T_{\phi l}$ are not symmetric, but for several separate
reasons (different ones in the semiclassical and harmonic regimes)
these functions, for the most part, are very nearly symmetric.
Actually, if the Born-Oppenheimer corrections to the decoherence
parameters, derived in Sec. \ref{bo}, were left out, $T_{1l}$ {\em
would} be exactly symmetric. \item The {\em s} curves ($T_{1s}$
and $T_{\phi s}$) are very different from the {\em l} curves
($T_{1l}$ and $T_{\phi l}$). This perhaps surprising result is
explained by the fact that the {\em s} functions have exactly no
contribution from the longitudinal term in the matrix elements
(first term in Eq. (\ref{fastslow})). The longitudinal term
usually dominates the transverse term (second term in Eq.
(\ref{fastslow})) when it is present, as it is for the {\em l}
functions.  As we will see, this makes the character of these
curves very different from one another.
\end{itemize}
\section{review of semiclassical analysis}
\label{semiclassical}

\begin{figure}[!t]
\includegraphics[width=8cm,trim=0 0 0 0]{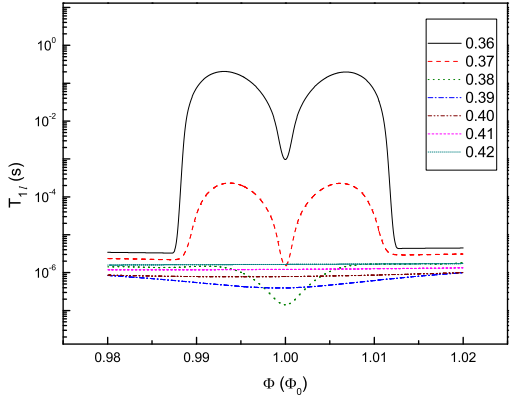}
\caption{The relaxation time $T_{1l}$ associated with the
dissipation source $Z_2$. The $T_{1l}$ plots are presented as a
function of changes in the small- and large-loop fluxes, $\Phi_c$
and $\Phi$ respectively.  As with $T_{1s}$, $T_{1l}$ has two
distinct regions, the ``semiclassical"
($\Phi_c\lesssim0.39\Phi_0$) and the ``harmonic"
($\Phi_c\gtrsim0.39\Phi_0$). The approximate symmetry around
$\Phi=\Phi_0$ arises from the dominant $\|$ contributions to the
matrix element Eq. (\ref{t1}).}
\label{t1l}
\end{figure}

As in BKD, we assume that the potential $U'({\bf f})$ describes a
double well with ``left'' and ``right'' minima at
\begin{eqnarray}
&&{\bf f}_L=\{{\rm f}_{\|L},{\bf f}_\perp^{min}({\rm f}_{\|L})\},\\
&&{\bf f}_R=\{{\rm f}_{\|R},{\bf f}_\perp^{min}({\rm f}_{\|R})\}.
\end{eqnarray}
Then, the semiclassical approximation amounts to assuming that the
left and right single-well ground states $|L\rangle$ and
$|R\rangle$ centered  at ${\bf f}_{L,R}$ are localized orbitals,
having amplitude that vanishes very rapidly away from these
minima. Then the two lowest eigenstates can approximately be
written as the symmetric and antisymmetric combinations of
$|R\rangle$ and $|L\rangle$,
\begin{eqnarray}
  |0\rangle  &=&  \frac{1}{\sqrt{2}}\left(\sqrt{1+\frac{\epsilon}{\omega_{01}}}\,|L\rangle + \sqrt{1-\frac{\epsilon}{\omega_{01}}}\,|R\rangle\right),\label{TL0}\\
  |1\rangle  &=&  \frac{1}{\sqrt{2}}\left(\sqrt{1-\frac{\epsilon}{\omega_{01}}}\,|L\rangle - \sqrt{1+\frac{\epsilon}{\omega_{01}}}\,|R\rangle\right),\label{TL1}
\end{eqnarray}
where $\omega_{01}=\sqrt{\Delta^2 +\epsilon^2}$, $\epsilon =
\langle L|{\cal H}_S|L\rangle - \langle R|{\cal H}_S|R\rangle$ is
the asymmetry of the double well, and $\Delta = \langle L|{\cal
H}_S|R\rangle$ is the tunneling amplitude between the two wells.
$\Delta$ increases almost exponentially with $\Phi_c$ as expected
in a WKB picture, see Fig. \ref{f01inset}.  Since $|L\rangle$ and
$|R\rangle$ are localized orbitals, we approximate the matrix
elements (see Eq. (\ref{fastslow})):
\begin{equation}
  \langle L|{\bf v}\cdot{\bf f}|R \rangle \approx  0,\quad
  \langle L|{\bf v}\cdot{\bf f}|L \rangle \approx  {\bf v}\cdot{\bf f}_L,\quad
  \langle R|{\bf v}\cdot{\bf f}|R \rangle \approx  {\bf v}\cdot{\bf f}_R.\label{SC-matrix}
\end{equation}
From Eqs.~(\ref{TL0})--(\ref{SC-matrix}) the eigenstate matrix
elements are
\begin{eqnarray}
  \langle 0|{\bf v}\cdot{\bf f}|1\rangle &\approx & \frac{1}{2}
  \frac{\Delta}{\omega_{01}} {\bf v}\cdot\Delta{\bf f},\label{t1}\\
  \langle 0|{\bf v}\cdot{\bf f}|0\rangle-\langle 1|{\bf v}\cdot{\bf f}|1\rangle
  &\approx & \frac{\epsilon}{\omega_{01}} {\bf v}\cdot\Delta{\bf f},\label{t2}
\end{eqnarray}
where $\Delta{\bf f} = {\bf f}_L-{\bf f}_R$.  These formulas will
be applied in different ways to explain the four quantities in
Figs. \ref{t1s}-\ref{tphl}.

In this semiclassical approximation with localized states, the
relaxation and decoherence times both diverge if $\Delta{\bf f}$ can
be made orthogonal to ${\bf v}$.  For a symmetric double well
($\epsilon =0$), $T_\phi\rightarrow\infty$ for all $\Delta {\bf f}$.

\section{$T_{1s}$}
\label{t1ssection}

\begin{figure}[!t]
\includegraphics[width=8cm,trim=0 0 0 0]{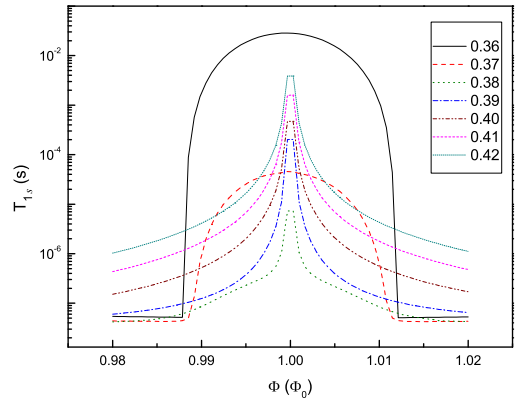}
\caption{The relaxation time $T_{1s}$ associated with the
dissipation source $Z_1$. The $T_{1s}$ plots are presented as a
function of changes in the small- and large-loop fluxes, $\Phi_c$
and $\Phi$ respectively. As a consequence of a high potential
barrier, up to $\Phi_c\approx 0.39$ the data can be well described
by a semiclassical model, see  Eq. (\ref{semiclassicalt1s}).  For
$\Phi_c\gtrsim0.39\Phi_0$, the behavior is nearly that in a
harmonic potential. The mathematical symmetry seen around
$\Phi=\Phi_0$ occurs because of the bilateral symmetry of the
qubit. The break around $\delta=\pm 0.01$
($\Phi=(1+\delta)\Phi_0$) indicates that for larger $\delta$ the
lowest two eigenstates are both located in one well}
\label{t1s}
\end{figure}

We see in Fig. \ref{t1s} that as $\Phi_c$ increases, $T_{1s}$
initially is almost constant in $\epsilon$ and decreasing
exponentially with $\Phi_c$; this behavior changes fairly abruptly
to one which is exponentially {\em increasing} in $\Phi_c$, with a
sharp maximum at $\epsilon=0$.

The initial behavior is explained by the semiclassical theory.  We
must specialize the semiclassical theory to a fact that is special
to the circuit coupling to the small loop: as a consequence of the
bilateral symmetry of the structure, the ``naive" ~longitudinal
contribution to the matrix elements vanishes, i.e.,
\begin{equation}
{\rm v}_\|=0.
\end{equation}
With this, we can specialize the $T_1$ matrix element Eq.
(\ref{t1}) thus:
\begin{equation}
|\langle 0|{\bf v}\cdot{\bf f}|1\rangle|^2 \approx  \frac{1}{4}
\frac{\Delta^2}{\omega_{01}^2} [{\bf v_\perp}\cdot
({\bf f}_\perp({\rm f}_\|^L)-{\bf f}_\perp({\rm f}_\|^R))]^2.
\end{equation}
We find that, again as a consequence of symmetry, the function
${\bf f}_\perp({\rm f}_\|)$ has a special form: at the
symmetric-well point, it is an even function of ${\rm f}_\|$
(assuming the origin is centered at the midpoint between the two
wells); in addition, this symmetry is broken continuously as
$\epsilon$ is made nonzero.  This can be summarized by writing the
start of the Taylor series for ${\bf v}\cdot{\bf f}_\perp({\rm
f}_\|)$:
\begin{equation}
{\bf v}\cdot{\bf f}_\perp({\rm f}_\|)=a
{\rm f}_\|^2+b\epsilon{\rm f}_\|+...\, .\label{s2222}
\end{equation}
Plugging this in and using ${\rm f}_\|^R=-{\rm f}_\|^L$ for
$\epsilon=0$ gives
\begin{equation}
|\langle 0|{\bf v}\cdot{\bf f}|1\rangle|^2 \approx  \frac{1}{4}
\frac{\Delta^2}{\omega_{01}^2}\times
b^2\epsilon^2({\rm f}_\|^R)^2,\label{t555}
\end{equation}
\begin{equation}
T_{1s}\propto{\omega_{01}^2\over\Delta^2\epsilon^2}={1\over\Delta^2}+{1\over\epsilon^2}.
\label{semiclassicalt1s}
\end{equation}
This simple functional form fits the curves in Fig. \ref{t1s} very
well for $\Phi_c=0.36-0.39\Phi_0$.

For larger $\Phi_c$ the trend of $T_{1s}$ is explained by the
observation that around $\Phi_c=0.39\Phi_0$, the barrier
disappears and the single minimum rapidly approaches being an
ideal harmonic potential.  If the potential were exactly harmonic,
with its minimum-curvature direction pointed in the $\|$
direction, then $T_1$ would diverge.  The exponential growth of
$T_{1s}$ in this regime reflects this approach to harmonicity.  At
all values of $\Phi_c$ it remains true that for $\Phi=\Phi_0$,
$T_{1s}$ is divergent, and the lineshapes around $\Phi=\Phi_0$
reflect this.

\section{$T_{1l}$}
\label{t1lsection}

$T_{1l}$ also has two distinct regions, the ``semiclassical" and
the ``harmonic".  In both regions, the longitudinal contributions
to the matrix element dominate.  This means that symmetry-breaking
contributions (for $\Phi=(1\pm\delta)\Phi_0$) remain very small in
all regimes (this is untrue for $T_{\phi l}$).

The semiclassical prediction for $T_{1l}$ is
\begin{equation}
T_{1l}^{-1}\propto|\langle 0|{\bf v}\cdot{\bf f}|1\rangle|^2 \approx
\frac{1}{4} \frac{\Delta^2}{\omega_{01}^2} [{\rm v}_\|
({\rm f}_\|^L-{\rm f}_\|^R)]^2.
\end{equation}
Since ${\rm f}_\|^{L,R}$ is slowly varying with $\Phi$ and
$\Phi_c$, we have
\begin{equation}
T_{1l}\propto 1+{\epsilon^2\over\Delta^2}.
\end{equation}
This equation predicts a $T_{1l}$ which is exponentially
decreasing overall, with a deep minimum at $\Phi=\Phi_0$, as seen
in the figure.
\begin{figure}[!t]
\includegraphics[width=8cm,trim=0 0 0 0]{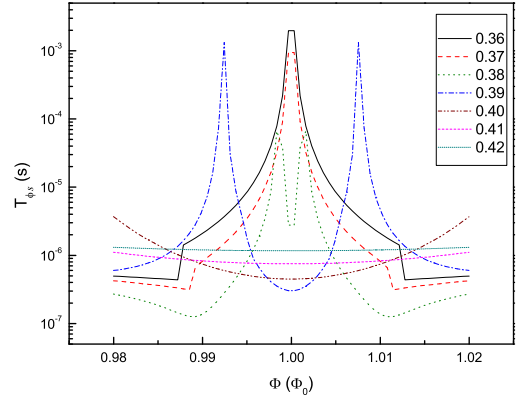}
\caption{The dephasing time associated with the dissipation source
$Z_1$. The $T_{\phi s}$ plots are presented as a function of
changes in the small- and large-loop fluxes, $\Phi_c$ and $\Phi$
respectively.  The two-peak structure is understood as a
manifestation of non-equal weight of the states $|0\rangle$ and
$|1\rangle$ in the matrix elements Eq. (\ref{t2}).  The
mathematical symmetry seen around $\Phi=\Phi_0$ occurs because of
the bilateral symmetry of the qubit.}
\label{tphs}
\end{figure}When the potential becomes harmonic, then $T_{1l}$ should approach
a constant almost independent of $\Phi$, since the harmonic
oscillator wavefunctions are only shifted by the force
proportional to $\Phi-\Phi_0$; the matrix element is independent
of this force. $T_{1l}$ is seen to slowly vary with $\Phi_c$: the
variation that is seen presumably reflects the small increase in
the harmonic frequency as $\Phi_c$ increases.

\section{$T_{\phi s}$}
\label{tphssection}

The semiclassical approximation follows the same development as
for $T_{1s}$, with the result (see (\ref{t555}))
\begin{equation}|\langle 0|{\bf v}\cdot{\bf f}|0\rangle-\langle 1|{\bf
v}\cdot{\bf f}|1\rangle|^2 \approx
\frac{\epsilon^2}{\omega_{01}^2}\times
b^2\epsilon^2({\rm f}_\|^R)^2\label{t777}
\end{equation}
\begin{equation}
T_{\phi
s}\propto{\omega_{01}^2\over\epsilon^4}={1\over\epsilon^2}+{\Delta^2\over\epsilon^4}
\end{equation}
This last equation predicts a strongly diverging $T_{\phi s}$ with
not very much $\Delta$ (i.e., $\Phi_c$) dependence, as is seen
initially in Fig. \ref{tphs}.

There is a fairly rapid departure from the semiclassical
prediction for $T_{\phi s}$ in that the divergence at
$\Phi=\Phi_0$ splits (symmetrically, as discussed above) into two
which rapidly move away from the center.  This is explained by the
fact that, once the wavefunctions become somewhat delocalized, the
difference between the 0 and 1 matrix elements in (\ref{t777})
becomes nonzero, pushing the divergence away from the symmetric
point.  This difference becomes nonzero because the 0 state (the
symmetric state at $\Phi=\Phi_0$) has more amplitude between the
two minima than the 1 (antisymmetric) state.  Thus, when weighted
by the ${\bf v}\cdot{\bf f}_\perp({\rm f}_\|)$ function (recall
Eq. (\ref{s2222})), the 0 and 1 matrix elements are (initially)
slightly different.

In the harmonic limit, $T_{\phi s}$ should diverge, just as
$T_{1s}$ does.  Comparing Figs. \ref{t1s}-\ref{tphs}, though,
shows that the details of this divergence are rather different. It
is evident that as this limit is approached, $T_{\phi l}$ is
dominated by the remaining differences of the 0 and 1 matrix
elements just discussed, which become nearly $\Phi$ independent as
the 0 and 1 wavefunctions become more harmonic.

\section{$T_{\phi l}$}
\label{tphlsection}

\begin{figure}[!t]
\includegraphics[width=8cm,trim=0 0 0 0]{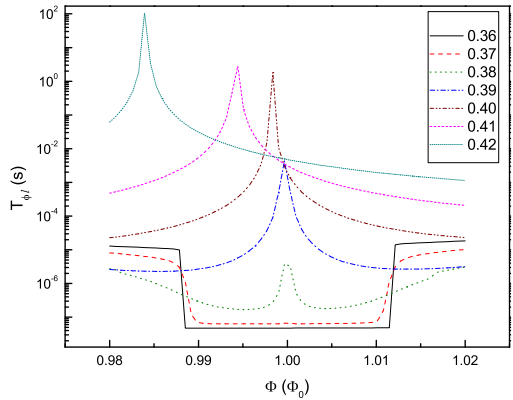}
\caption{The dephasing time associated with the dissipation source
$Z_2$. The $T_{\phi l}$ plots are presented as a function of
changes in the small- and large-loop fluxes, $\Phi_c$ and $\Phi$
respectively.  The significant breaking of symmetry around the S
line occurs due the transverse contributions to the matrix
elements Eq. (\ref{t2}), and is associated with the unsymmetrical
way that this source couples to the qubit.}
\label{tphl}
\end{figure}

$T_{\phi l}$  is shown in Fig. \ref{tphl}.  In the
semiclassical regime this should be
\begin{equation}
T_{\phi l}^{-1}\propto|\langle 0|{\bf v}\cdot{\bf
f}|0\rangle-\langle 1|{\bf v}\cdot{\bf f}|1\rangle|^2 \approx
\frac{\epsilon^2}{\omega_{01}^2} [{\rm v}_\| ({\rm f}_\|^L-{\rm
f}_\|^R)]^2,
\end{equation}
So
\begin{equation}
T_{\phi l}\propto\frac{\epsilon^2+\Delta^2}{\epsilon^2}.
\end{equation}
For small $\Delta$ this predicts, as seen, an almost
$\Phi_c-$independent behavior, with a very weak divergence at
$\Phi=\Phi_0$.  As $\Delta$ increases, the divergence gets
stronger and $T_{\phi l}$ begins to increase overall.

In the harmonic limit, again, $T_{\phi l}$ should diverge.  But
as the $\|$ contributions disappear, eventually the transverse
contributions to the matrix elements begin to be important.  These
explicitly break the symmetry, as can easily be seen as the
shifting of the divergence point in the last few $T_{\phi l}$
curves.

However, this asymmetry is unlikely to be noticeable
experimentally.  Recall that the physical $T_\phi$ and $T_1$ are
(approximately \cite{BB}) given by summing the $s$ and $l$ rates.
The strong asymmetry in $T_{\phi l}$ occurs only when its
contribution to the rate is very small, and the symmetric
$T_{\phi s}$ will dominate.

\section{Discussion and Conclusions}
\label{conclusions}

\begin{figure}[!tb]
\includegraphics[width=8cm,trim=0 0 0 0]{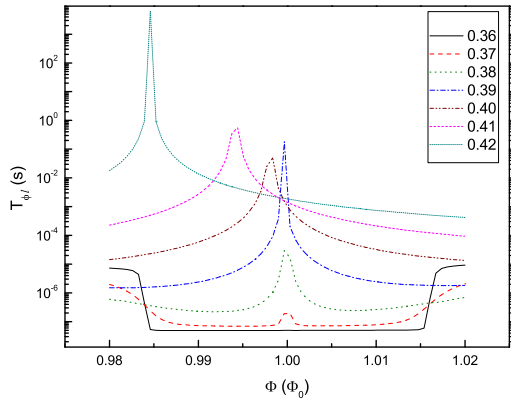}
\vspace*{-0.25cm}\caption{$T_{\phi l} $ for the case without stray capacitances
($C_4=C_5=0)$. We see the same qualitative behavior as that
observed in Fig. \ref{tphs}.}
\label{tphl3}
\end{figure}

We conclude with a discussion of the effect of the presence of
stray capacitances, and on the overall implication of our results
on decoherence parameters for experiments on the gradiometer
qubit.
\begin{figure}[!t]
\includegraphics[width=8cm,trim=0 0 0 0]{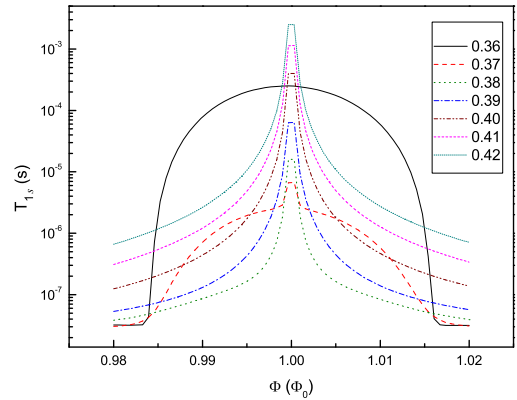}
\caption{$T_{1s} $ for the case without stray capacitances
($C_4=C_5=0)$. We see the same qualitative behavior compared with
that observed in Fig. \ref{t1s}.}
\label{t1s3}
\end{figure}
Figs. \ref{t1s3} and \ref{tphl3} show the results for $T_{1s}$ and
$T_{\phi l}$ for the gradiometer qubit with zero assumed stray
capacitances.  Graphically, the results are apparently only
slightly changed.  This is somewhat an impression created by the
log scale; a closer examination of the $T_1$ result shows that the
presence of stray capacitances actually {\em improves} the
relaxation time (i.e., makes it longer) by about a factor of 10 in
the double-well region, while leaving it more or less unchanged in
the harmonic, single-well region.

\begin{figure}[!t]
\includegraphics[width=8cm,trim=0 0 0 0]{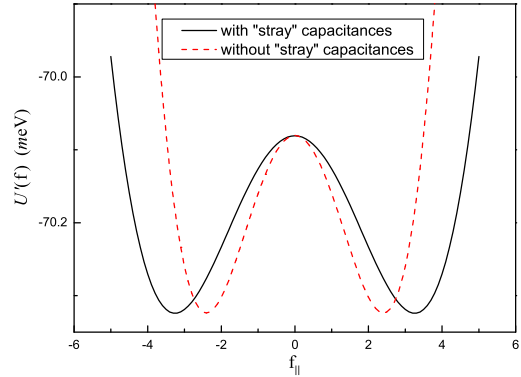}
\caption{Potential $U'({\bf f})$, Eq. \ref{rescaledpotential},
along the ``slow" coordinate ${\rm f}_\|$ for the cases with
(black solid line) and without (red dashed line) stray
capacitances.  The same effective capacitance $c$ is used for
both.  It is observed that the height of barrier is almost the
same for both cases (the difference is less than $10^{-6}\%$)
while the distance between the minima changes appreciably. Because
of these facts one might expect that the tunnelling rate should be
higher for the case without ``stray" capacitances, and thus the
$T_1$ should be shorter. This is seen in our calculations
--- compare Figs. \ref{t1l} and \ref{t1s3}. The external
fluxes used for these plots are $\Phi_c=0.36\Phi_0$ and
$\Phi=\Phi_0$.}
\label{potential53}
\end{figure}

\begin{figure}[!t]
\includegraphics[width=8cm,trim=0 0 0 0]{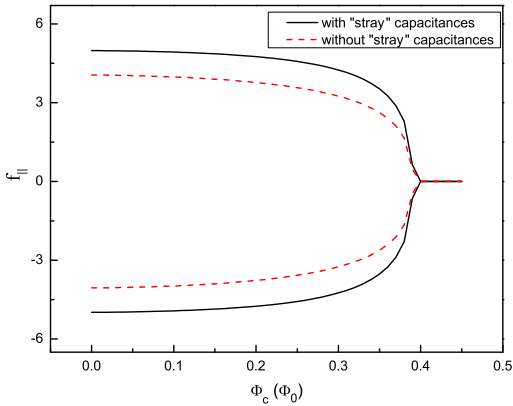}
\caption{The ``slow" coordinate f$_\|$ minima, for the relevant
minima (L$_i$-R$_i$), as a function of $\Phi_c$ along the S line.
We compare the evolution for the cases with (black solid line) and
without (red dashed line) stray capacitances. }
\label{phaseparallel}
\end{figure}

Figs. \ref{potential53} and \ref{phaseparallel} provide some
explanation for this observation.  We see that the double-well
potential profile is very similar in the two cases, with the well
depths being virtually the same. However, the presence of the
strays pushes apart the rescaled distance between the two minima.
This diminishes the tunneling between the two wells, and, not
surprisingly therefore, lengthens the relaxation time to go from
one well to the other. In Fig. \ref{phaseparallel} we see that
this effect persists right up to the point where the two minima
merge at around $\Phi_c=0.39\Phi_0$.

\begin{figure}[!t]
\includegraphics[width=8cm,trim=0 0 0 0]{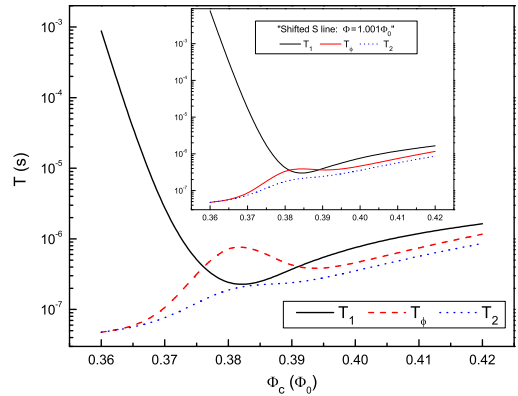}
\caption{The total relaxation, dephasing and decoherence times
($T_1$, $T_\phi$ and $T_2$, respectively) along the S line.  We
can see that $T_\phi$ ($T_1$) strongly increases (decreases) as
a function of $\Phi_c$. These facts cause there to be a window of
desirable operating parameters for the qubit.}
\label{ttotalsline}
\end{figure}

Finally, Fig. \ref{ttotalsline} gives perhaps the most
experimentally relevant summary of our results for the realistic
gradiometer qubit parameters (with stray capacitances).  During
qubit operation, it is envisioned \cite{Kochetal} that the qubit
will be initialized at small control flux, and will then be pulsed
rapidly up to high control flux; above $\Phi_c=0.38-0.39\Phi_0$,
we expect the coherence of the qubit to be protected by an
oscillator stabilization not discussed here.

The preferred initialization point is at a value of $\Phi_c$ well
below $0.36\Phi_0$.  We see that here we have the right conditions
for initialization of the qubit, in that the $T_1$ time will be
very long -- the figure shows it increasing exponentially as
$\Phi_c$ is decreased.  This is a simple reflection of the very
large barrier height in this region.  $T_\phi$, and therefore
$T_2$ are very small in this region, but this is not harmful
during initialization.  When $\Phi_c$ is pulsed upwards, quantum
dynamics turn on in a region around $\Phi_c=0.36-0.37\Phi_0$,
where the 0-1 frequency is increasing exponentially through the
100MHz range. This is the relevant frequency because it is roughly
the inverse of the anticipated rise time of the pulse in this
region, 1-10nsec; this is referred to as the ``portal" region in
\cite{Kochetal}.

Thus, the crucial region for the operation of the qubit is in the
range of $\Phi_c=0.37-0.38\Phi_0$.  It is evidently a very
perilous region for the qubit: $T_1$ is plunging downward,
dropping to around 200nsec, and $T_2$ is increasing from its very
small values in the initialization region, but does not rise far
beyond 100nsec in this region.  Since we expect \cite{Kochetal} to
pulse through this region in under 10nsec, these times are
acceptable for qubit operation; but we see that the qubit could
not function if it were held in this region of $\Phi_c$ without
any other protective mechanism for a long time.  Also, we must
beware of other effects, such as higher effective temperatures or
stronger mutual inductance couplings, that would make these times
even worse.  We believe that there have been occasions when the
conditions of the experiment were worse by a factor of 10 or more;
in this case, the qubit's coherence is not likely to survive even
a 10nsec traversal of this region of parameter space.

Fig. \ref{ttotalsline} is obtained by adding the inverse relation
times arising from the small-loop and large-loop circuits (we
ignore the small nonadditive effects explored in \cite{BB}).
$T_1$ is, though the whole region of interest, dominated by
$T_{1l}$.  For small $\Phi_c$, $T_\phi$ is also dominated by
the large-loop circuit; however, for $\Phi_c>0.38\Phi_0$ the
small-loop dephasing becomes dominant.  So, we see that the
analysis of both circuits is experimentally relevant.

Finally, the inset to Fig. \ref{ttotalsline} shows the effects of
imperfections in the setting of $\Phi$, which would put the system
off the S line.  We see that, even for departures of $1m\Phi_0$,
there are noticeable changes in the decoherence parameters.  $T_1$
is actually increased, reflecting the fact that $T_{1l}$ has a
minimum at the S line.  But the system's sensitivity to phase
fluctuations increases -- $T_\phi$ is smaller off the S line.
While the changes seen $1m\Phi_0$ from the S line are not
dramatic, these results indicate that much larger departures
should be treated with caution, and subject to a full analysis.
A concluding word:  The results reported here have been a useful
guide to experiment, but they have shown their greatest worth when
they are part of the iterative design process itself.  Thus, even
the rather complete snapshot given here does not do justice to the
full role that this analysis has had in the process of perfecting
the gradiometer qubit.  The work has already passed on to further
questions not touched on here, such as the role of additional
harmonic oscillator circuits in modifying the decoherence
parameters \cite{Kochetal, Kochnew}, and the problems created by
introducing qubit-qubit coupling.  An intimate relationship
between theory and experiment will continue to be crucial in the
continuing development of this qubit technology.

\section*{Acknowledgments}

DPDV is supported in part by the NSA and ARDA through ARO contract
number W911NF-04-C-0098.  FB is supported by Funda\c{c}{\~a}o de Amparo {\`a} Pesquisa do Estado de S{\~a}o Paulo (FAPESP). RHK thanks the support of DARPA under contract
MDA972-01-C-0052.
\section*{Appendix: Circuit Theory}
\label{appendix}

We make extensive use here of the systematic analysis of flux
qubits initiated in BKD.  For completeness, we review the
formalism presented there, both the network graph theory and the
Caldeira-Leggett analysis.  We have added some extensions to this
theory, which we separate out and present in a separate subsection
of this appendix.

\subsection{Review of BKD}
\label{reviewbkd}

This subsection is a streamlined summary of the results presented
in \cite{BKD}.

An {\em oriented graph} ${\cal G}=({\cal N},{\cal B})$ consists of
$N$ nodes ${\cal N}=\{n_1,\ldots, n_N\}$ and $B$ branches ${\cal
B}=\{b_1,\ldots, b_B\}$. In circuit analysis, a {\em branch}
$b_i=(n_{a(i)},n_{b(i)})$ represents a two-terminal element
(resistor, capacitor, inductor, current or voltage source, etc.),
connecting its beginning node $n_{a(i)}$ to its ending node
$n_{b(i)}$.  A {\em loop} in ${\cal G}$ is a connected subgraph of
${\cal G}$ in which all nodes have degree two (the degree of a node $n  \in {\cal N}$ is the number of branches containing $n$).  For each connected
subgraph we choose a {\em tree} ${\cal T}_i$, i.e.\ a connected
subgraph of ${\cal G}_i$ which contains all its nodes and has no
loops. The branches that do not belong to the tree are called {\em
chords}.  The {\em fundamental loops} ${\cal F}_i$ of a subgraph
${\cal G}_i$ are defined as the set of loops in ${\cal G}_i$ which
contain exactly one chord $f_i\in {\cal G}_i\backslash {\cal
T}_i$.

A complete description of the topology of the network is provided
by the {\em fundamental loop matrix}, defined as
\begin{equation}
  {\bf F}^{(L)}_{ij} = \left\{\begin{array}{r l}
      1, & \mbox{if $b_j\in{\cal F}_i$ (same direction as $f_i$)},\\
     -1, & \mbox{if $b_j\in{\cal F}_i$ (direction opposite to $f_i$)},\\
      0, & \mbox{if $b_j\not\in{\cal F}_i$},
\end{array}\right.
\end{equation}
where $i=1,\ldots F$ and $j=1,\ldots, B$.  By labeling the
branches of the graph ${\cal G}$ such that the first $N-P$
branches belong to the tree ${\cal T}$, where $P$ is the number of
disjoint connected subgraphs of ${\cal G}$, we obtain
\begin{equation}
  \label{QF}
  {\bf F}^{(L)} = \left( -\bF^T \, |\, \openone\right),
  \label{loopmatrix}
\end{equation}
where ${\bf F}$, the {\em loop matrix}, is an $(N-P)\times(B-N+P)$
matrix.

The state of an electric circuit described by a network graph can
be defined by the branch currents ${\bf I}=(I_1,\ldots I_B)$,
where $I_i$ denotes the electric current flowing in branch $b_i$,
and the branch voltages ${\bf V}=(V_1,\ldots V_B)$, where $V_i$
denotes the voltage drop across the branch $b_i$.  If we divide
the branch currents and voltages into a tree and a chord part,
\begin{eqnarray}
  \label{tree-chord-separation}
  {\bf I} &=& ({\bf I}_{\rm tr}, {\bf I}_{\rm ch}),\\
  {\bf V} &=& ({\bf V}_{\rm tr}, {\bf V}_{\rm ch}),
\end{eqnarray}
Then the Kirchhoff laws can be stated very succinctly and
universally:
\begin{eqnarray}
  {\bf F}  {\bf I}_{\rm ch} &=& - {\bf I}_{\rm tr},\label{IVct-1}\\
  {\bf F}^T{\bf V}_{\rm tr} &=&   {\bf V}_{\rm ch} - {\bf \dot\Phi}_x.\label{IVct-2}
\end{eqnarray}
Here ${\bf \Phi}_x$ are the external magnetic fluxes threading the
loops.

To write the Hamiltonian of the electrical circuit, we must
further distinguish the different types of electrical circuit
elements in the graph.  We write
\begin{eqnarray}
  \label{Fsplit}
  {\bf F} = \left(\begin{array}{c c c c c}
      {\bf F}_{CJ} & {\bf F}_{CL} & {\bf F}_{CR} & {\bf F}_{CZ} & {\bf F}_{CB} \\
      {\bf F}_{KJ} & {\bf F}_{KL} & {\bf F}_{KR} & {\bf F}_{KZ} & {\bf F}_{KB}
\end{array}\right).
\end{eqnarray}
The sub-matrices ${\bf F}_{XY}$ will be called {\em loop
sub-matrices}.  The different chord labels are for Josephson
junctions (J), linear inductors (L), shunt resistors (R) and other
external impedances (Z), and bias current sources (B).  Without
loss of generality the capacitors (C) can all be taken as tree
branches.  The tree inductors are labeled (K).  We note here that
in our formalism all capacitors should be considered to be in
parallel with a Josephson junction, even if it is one with zero
critical current.

Finally, to fully define the problem, the electrical
characteristics of each branch type should be defined.  The
current-voltage relations for the various types of branches are
\begin{eqnarray}
  {\bf I}_J  &=&  {\bf I}_{\rm c} \,\mbox{\boldmath $\sin$} \bphi ,\label{CVR-J}\\
  {\bf Q}_C  &=&  {\bf C}{\bf V}_C,\label{CVR-C}\\
  {\bf V}_R  &=&  {\bf R}{\bf I}_R\label{CVR-R},\\
  {\bf V}_Z(\omega)  &=&  {\bf Z}(\omega){\bf
  I}_Z(\omega)\label{CVR-Z},\\
  \label{inductance-1}
   \left(\begin{array}{c}{\bf \Phi}_L\\ {\bf
   \Phi}_K\end{array}\right)&
  = & \left(\begin{array}{l l}{\bf L}        & {\bf L}_{LK}\\
                           {\bf L}_{LK}^T & {\bf L}_{K} \end{array}\right)
   \left(\begin{array}{c}{\bf I}_L\\ {\bf I}_K\end{array}\right).
\end{eqnarray}
Here the diagonal matrix ${\bf I}_{\rm c}$ contains the critical
currents $I_{{\rm c},i}$ of the junctions on its diagonal, and
$\mbox{\boldmath $\sin$}\bphi$ is the vector
$(\sin\varphi_1,\sin\varphi_2,\ldots,\sin\varphi_{N_J})$.
Eq.~(\ref{CVR-C}) describes the (linear) capacitors (${\bf C}$ is
the capacitance matrix), and the junction shunt resistors are
described by Eq.~(\ref{CVR-R}) where $R$ is the (diagonal and
real) shunt resistance matrix. The external impedances are
described by the relation Eq.~(\ref{CVR-Z}) between the Fourier
transforms of the current and voltage,  where ${\bf Z}(\omega)$ is
the impedance matrix. The external impedances can also defined in
the time domain,
\begin{equation}
  \label{CVR-Z-time}
  {\bf V}_Z(t)=\int_{-\infty}^t {\bf Z}(t-\tau){\bf I}_Z(\tau) d\tau \equiv ({\bf Z}*{\bf I}_Z)(t),
\end{equation}
where the convolution is defined as
\begin{equation}
  \label{convolution}
  ({\bf f} * {\bf g}) (t) = \int_{-\infty}^t {\bf f}(t - \tau) {\bf g}(\tau) d\tau .
\end{equation}
Causality allows the response function to be nonzero only for
positive times, ${\bf Z}(t)=0$ for $t<0$. In frequency space, the
replacement $\omega\rightarrow \omega + i\epsilon$ with
$\epsilon>0$ guarantees convergence of the Fourier transform
\footnote{We choose the Fourier transform such that it yields the
impedance $Z(\omega)=+i\omega L$ for an inductor (inductance
$L$).}
\begin{equation}
  \label{Z-FT}
  {\bf Z}(\omega) = \int_{-\infty}^\infty {\bf Z}(t)e^{i\omega t}dt
                  = \int_0^\infty {\bf Z}(t)e^{i\omega t}dt .
\end{equation}
In our formalism it is necessary to distinguish chord from tree
inductors, so the inductance matrix must be written in block form
shown in Eq. (\ref{inductance-1}).

With all these definitions, a universal equation of motion for the
electric circuit reads
\begin{widetext}\beq \bC{\ddot\bphi}=-\bL_J^{-1}\bsin\bphi-\bM_0\bphi
-{2\pi\over\Phi_0}(\bar{\bN}*\bPhi_x)(t)
-{2\pi\over\Phi_0}(\bar{\bS}*\bI_B)(t)-\bR^{-1}{\dot\bphi}-\bM_d*\bphi.\label{univ}
\eeq \end{widetext}This equation as presented is a slight extension of BKD, in
that $\bPhi_x$ and $\bI_B$ are allowed to be time dependent.  If
they are time independent, then the expressions for the
coefficients of this equation of motion are:

The coefficients of this equation are as follows
\begin{equation}
  \label{Mcircuit}
  \bM_d(\omega) = \bar{\bf m}\bar{\bf L}_Z(\omega)^{-1}\bar{\bf m}^T,
\end{equation}
\beq \bM_0=\bar{\bN}(\omega=0)\bF_{\text{CL}}^T \label{M0},\eeq \beq
\bar{\bN}(\omega=0)=\bF_{\text{CL}}\tilde{\bL}_{\text{L}}^{-1}\bar{\bL}\bL_{\text{LL}}^{-1}. \label{Nbar}
\eeq
\begin{equation}
  \label{Mcircuit}
  \bM_d(\omega) = \bar{\bf m}\bar{\bf L}_Z(\omega)^{-1}\bar{\bf m}^T,
\end{equation}
\beq
\bar{\bS}(\omega=0)=\bF_{\text{CB}}-\bF_{\text{CL}}(\bL_{\text{LL}}^{-1})^T
\bar{\bF}_{\text{KL}}^T\tilde{\bL}_{\text{K}}^T\bF_{\text{KB}},
\eeq \beq \bar{\bf
m}=\bF_{\text{CZ}}-\bF_{\text{CL}}(\bL_{\text{LL}}^{-1})^T
\bar{\bF}_{\text{KL}}^T\tilde{\bL}_{\text{K}}^T\bF_{\text{KZ}}.
\eeq With the definitions, as given in \cite{BKD}: \beq
\bar{\bL}_{\text{Z}}=\bL_{\text{ZZ}}-\bL_{\text{ZL}}\bL_{\text{LL}}^{-1}\bL_{\text{LZ}},
\eeq \beq
\bar{\bL}_{\text{L}}(\omega)=\bL_{\text{LL}}-\bL_{\text{LZ}}
\bL_{\text{ZZ}}(\omega)^{-1}\bL_{\text{ZL}}, \eeq \beq
\bL_{\text{ZL}}=\bF_{\text{KZ}}^T\tilde{\bL}_{\text{K}}\bar{\bF}_{\text{KL}},
\eeq \beq
\bL_{\text{LZ}}=\bF_{\text{KL}}^T\tilde{\bL}_{\text{K}}\bF_{\text{KZ}},
\eeq \beq
\bL_{\text{ZZ}}=\bL_{\text{Z}}+\bF_{\text{KZ}}^T\tilde{\bL}_{\text{K}}\bF_{\text{KZ}},
\eeq \beq
\bL_{\text{LL}}=\bar{\bL}+\bF_{\text{KL}}^T\tilde{\bL}_{\text{K}}\bar{\bF}_{\text{KL}},
\eeq \beq
\bar{\bF}_{\text{CY}}=\bF_{\text{CY}}+\bF_{\text{CL}}\bL^{-1}\bL_{\text{LK}}
\bar{\bL}_{\text{K}}^{-1}\tilde{\bL}_{\text{K}}\bF_{\text{KY}},\,\,\,Y=Z,B,
\eeq \beq
\bar{\bF}_{\text{KL}}=\bF_{\text{KL}}-\bL_{\text{K}}^{-1}\bL_{\text{LK}}^T,
\eeq \beq
\tilde{\bL}_{\text{L}}=\bar{\bL}\left(\bone_{\text{L}}+\bL^{-1}\bL_{\text{LK}}
\bar{\bL}_{\text{K}}^{-1}\tilde{\bL}_{\text{K}}\bar{\bF}_{\text{KL}}\right)^{-1},
\eeq \beq
\tilde{\bL}_{\text{K}}=\bar{\bL}_{\text{K}}\left(\bone_{\text{K}}-\bL_{\text{K}}
\bar{\bF}_{\text{KL}}\bL^{-1}\bL_{\text{LK}}\bar{\bL}_{\text{K}}^{-1}\right)^{-1},
\eeq \beq
\bar{\bL}_{\text{K}}=\bL_{\text{K}}-\bL_{\text{LK}}^T\bL^{-1}\bL_{\text{LK}},
\eeq \beq
\bar{\bL}=\bL-\bL_{\text{LK}}\bL_{\text{K}}^{-1}\bL_{\text{LK}}^T.
\eeq

The last two terms of Eq. (\ref{univ}) describe dissipation,
handled by the Caldeira-Leggett (see \cite{CG}), to be reviewed
shortly. The remaining terms are generated by a (time dependent)
system Hamiltonian: \beq \cH_S(t)={1\over
2}\bQ_C^T\bC^{-1}\bQ_C+\left({\Phi_0\over2\pi}\right)U(\bphi,t),\label{generalpotential}
\eeq
\bea
\lefteqn{\!\!\!\!\!\!\!U(\bphi,t)=-\sum_iL_{J;i}^{-1}\cos\varphi_i}\nonumber\\&&\!\!\!\!\!\!\!\!\!+{1\over
2}\bphi^T\bM_0\bphi+{2\pi\over\Phi_0}\bphi^T[(\bar{\bN}*\bPhi_x)(t)+
(\bar{\bS}*\bI_B)(t)]. \eea

Performing a standard Born-Markov approximation for the system
dynamics, one obtains predictions for the relaxation times of the
system:
\begin{eqnarray}
  \frac{1}{T_1} &=& 4|\langle 0|{\bf m}\cdot\bphi|1\rangle|^2 J(\omega_{01}) \coth\frac{\omega_{01}}{2k_B T}, \label{T1}\\
  \frac{1}{T_\phi} &=&  |\langle 0|{\bf m}\cdot\bphi|0\rangle-\langle 1|{\bf m}\cdot\bphi|1\rangle|^2 \left.\frac{J(\omega)}{\omega}\right|_{\omega\rightarrow 0} \!\!\!\!\!\!\!\!\! 2k_B T. \quad\quad\label{Tphi}
\end{eqnarray}

We consider the external impedances contributing to decoherence
one at a time (for an analysis of non-additive effects, see \cite{BB}).
${\bf M}_d (\omega)$, which determines the quantities in Eq.
(\ref{T1}, \ref{Tphi}), then has the form,
\begin{eqnarray}
   {\bf M}_d (\omega) &=& \mu K(\omega) {\bf m}{\bf m}^T,\label{Md-simple}\\
   K(\omega)          &=& \bar{\bf L}_Z^{-1}(\omega),\label{K-def}\\
   \mu                &=& |\bar{\bf m}|^2,\label{mu}\\
   {\bf m}            &=& \bar{\bf m}/\sqrt{\mu}=\bar{\bf m}/|\bar{\bf m}|,\label{m}
\end{eqnarray}
where $K(t)$ is a scalar real function, ${\bf m}$ is the
normalized vector parallel to $\bar{\bf m}$, and $\sqrt{\mu}$ is
the length of the vector $\bar{\bf m}$ ($\mu$ is the eigenvalue of
the rank 1 matrix $\bar{\bf m}\bar{\bf m}^T$).  Also\begin{equation}
  J(\omega) = - \mu \left(\frac{\Phi_0}{2\pi}\right)^2 {\rm Im} K(\omega).
  \label{JK}
\end{equation}
\subsection{New results}
\label{newresults}

The following expressions were not contained in BKD, and are new
to this paper.

The full frequency dependent expressions for $\bar\bS$ and
$\bar\bN$ are:
\begin{widetext}
\bea \bar{\bS}(\omega)=\bar{\bF}_{\text{CB}}-\left[
\bar{\bF}_{\text{CZ}}\bar{\bL}_{\text{Z}}^{-1}(\omega)\left(\bF_{\text{KZ}}^T-
\bL_{\text{ZL}}\bL_{LL}^{-1}\bF_{\text{KL}}^T\right)+\bF_{\text{CL}}\tilde{\bL}_{\text{L}}^{-1}
\bar{\bL}\bar{\bL}_{\text{L}}^{-1}\left(\bF_{\text{KL}}^T-
\bL_{\text{LZ}}\bL_{\text{ZZ}}^{-1}(\omega)\bF_{\text{KZ}}^T\right)\right]
\tilde{\bL}_{\text{K}}\bF_{\text{KB}},\eea \beq
\bar{\bN}(\omega)=\bF_{\text{CL}}\tilde{\bL}_{\text{L}}^{-1}\bL\bL_{\text{L}}^{-1}(\omega)-
\bar{\bF}_{\text{CZ}}\bar{\bL}_Z^{-1}(\omega)\bL_{\text{ZL}}\bL_{LL}^{-1}.
\eeq \end{widetext}In our previous work we assumed that $\bI_B$ and $\bPhi_x$
were time independent, so that only the $\omega\rightarrow 0$
limit of these expressions were presented.

A final result: it is amusing the write out the full expression
for $\bar{\bN}(\omega=0)$, from which $\bM_0$ is easily
constructed, in terms of the basic input matrices (loop matrices
{\bf F} and inductance matrices {\bf L}):
\begin{widetext} \bea \lefteqn{\bar{\bN}(\omega=0)=
\bF_{\text{CL}}\Bigg[\bone_{\text{L}}+\bL^{-1}\bL_{\text{LK}}\Big(\bL_{\text{K}}-\bL_{\text{LK}}^T\bL^{-1}\bL_{\text{LK}}
   \Big)^{-1}}\nonumber\\&&\!\!
   \left(\bone_{\text{K}}-\bL_{\text{K}}\left(\bF_{\text{KL}}-\bL_{\text{K}}^{-1}\bL_{\text{LK}}^T\right)\bL^{-1}
   \bL_{\text{LK}}\left(\bL_{\text{K}}-\bL_{\text{LK}}^T\bL^{-1}\bL_{\text{LK}}
   \right)^{-1}\right)^{-1}\bL_{\text{K}}\left(\bF_{\text{KL}}-\bL_{\text{K}}^{-1}\bL_{\text{LK}}^T\right)\Bigg]
   \\&&\!\!\!\!
   \left[\bL-\bL_{\text{LK}}\bL_{\text{K}}^{-1}\bL_{\text{LK}}^T+\bF_{\text{KL}}^T   \left(\bone_{\text{K}}-\bL_{\text{K}}\left(\bF_{\text{KL}}-\bL_{\text{K}}^{-1}\bL_{\text{LK}}^T\right)
   \bL^{-1}\bL_{\text{LK}}\left(\bL_{\text{K}}-\bL_{\text{LK}}^T\bL^{-1}
   \bL_{\text{LK}}\right)^{-1}\right)^{-1}\bL_{\text{K}}\left(\bF_{\text{KL}}-
   \bL_{\text{K}}^{-1}\bL_{\text{LK}}^T\right)\right]^{-1}.\nonumber
   \eea \end{widetext}
It is clear why it is more manageable to write this expression in
terms of intermediate quantities; one can see that it involves up to
four nested inverses.

\end{document}